\documentclass[11pt,a4paper]{article}

\usepackage[a4paper,margin=27mm]{geometry}
\usepackage[T1]{fontenc}
\usepackage[utf8]{inputenc}
\usepackage{lmodern}
\usepackage{textcomp}
\usepackage{pdfpages}
\usepackage{amsmath,amssymb}
\usepackage{microtype}
\usepackage{graphicx}
\graphicspath{{figures/}}
\usepackage{booktabs}
\usepackage{longtable}
\usepackage{rotating}
\usepackage[below]{placeins}   %

\usepackage{array}
\usepackage{calc}
\usepackage{xcolor}
\usepackage{caption}
\usepackage{enumitem}
\setlist{nosep,leftmargin=*}
\usepackage[round,authoryear,sort]{natbib}
\usepackage[hidelinks]{hyperref}
\usepackage{xurl}  %
\usepackage{setspace}
\newif\iffinal\finaltrue
\newif\ifdraftnotes\draftnotesfalse
\newcommand{\src}[1]{}\newcommand{\status}[1]{}\newcommand{\todo}[1]{}\newcommand{\pending}[1]{}

\title{Missed Fouls under Selective Review:\\[0.3em]
\large What the NBA's Last Two Minute Reports Identify}
\hypersetup{pdftitle={Missed Fouls under Selective Review: What the NBA's Last Two Minute Reports Identify},pdfauthor={Tianyu Guan and Yang Zhou}}
\author{Tianyu Guan\thanks{Department of Mathematics and Statistics, York University, Toronto, Canada, and TY Sports Analytics Inc., Canada. ORCID 0009-0006-7268-7966. Corresponding author: \href{mailto:tguan@yorku.ca}{tguan@yorku.ca}.}
\and Yang Zhou\thanks{TY Sports Analytics Inc., Canada. ORCID 0000-0001-9057-6713. \href{mailto:yangzhou@tysportsanalytics.com}{yangzhou@tysportsanalytics.com}.}}

\date{}

\begin{document}
\maketitle

\begin{abstract}
The NBA's Last Two Minute reports review every call in the closing minutes of close games but list only the non-calls their reviewers judge material, so missed fouls are seen selectively. In 2015-16, to our knowledge the only season with public player tracking, reports and tracking overlap on 223 games, with 170 listed missed calls. First, the report counts alone bound the share of true fouls called, as the league grades them: at most about 78~\% (0.777 {[}0.744, 0.806{]}), however complete the listing is (exploratory). Second, if every missed foul is listed, the propensity of a true foul to be called is identified. In the registered analysis it rises with a calling score from pre-whistle geometry (logit slope 0.390 {[}0.220, 0.577{]}), provided completeness does not fall and relative retention does not rise with the score; exploratory checks place part of the rise between foul types, and within types it is suggested, not established. Third, a pre-registered test found no strong separation of listed missed fouls from listed correct non-calls by that score (area under the ROC curve 0.469 {[}0.415, 0.526{]}); it concerns ranking among the plays the reviewers chose to list. Completeness and transport beyond the review scope are not identified and enter as bounds and tipping points.
\end{abstract}

\noindent\textbf{Keywords.} verification bias; partial identification; selective labels; player tracking; officiating; pre-registration.

\section{Introduction}\label{sec:1}

A foul in basketball is decided three times: on the floor, in real time and from one angle; after a close game, when the league reviews on video every call and the material non-calls of the closing minutes and publishes an assessment of each; and in advance, by the rulebook. An incorrect call can be seen and argued about, whereas a missed foul is invisible on the floor and in the play-by-play. We compare the floor with the league's review and use the rulebook only to define the rule predicates of Section~\ref{sec:5.1}.

We ask what an official review that covers only some decisions can tell us about them, with the missed call as the case. Calls are observed for every contact; correctness only for the contacts the league reviewed: \emph{all} calls, but only the non-calls its reviewers find material, and only inside a review scope (Section~\ref{sec:2.1}). A model trained on the \emph{listed} plays inherits the reviewers' selection of non-calls. The structure is outcome misclassification under selective verification, close to verification bias in diagnostic testing \citep{begg1983}, positive--unlabeled learning \citep{bekker2019} and selective labels \citep{lakkaraju2017}; the identification argument of Section~\ref{sec:4} is elementary in that literature. What is specific here is a review that lists every call, and the measured ways in which the review leaks into tracking-derived data (Section~\ref{sec:7.3}). We take the league's assessment as the outcome: a \emph{true foul} is a contact (a proximity segment of an offence--defence pair; Section~\ref{sec:5.1}) that the review graded, or would have graded, as a foul. Whether the assessments are right by the rule is a separate question (Section~\ref{sec:4.6}).

The paper makes three points. \emph{First, a bound from the report counts.} Because the completeness of the listing is at most 1, the reports' own counts bound the share of true fouls called, with no tracking, model or complete listing: at most about 78~\% in the review scope (Section~\ref{sec:3}). This is the worst-case bound of a review that verifies every positive decision \citep{kosinski2003b}; it is usable here because the league verifies every call. \emph{Second, what the review identifies.} Under \emph{completeness} (every uncalled true foul inside the review scope is listed as a missed call), the listed true fouls are all the true fouls, so the propensity of a true foul to be called is identified from them alone, whatever rule selects the listed non-fouls; because every call is listed, the prevalence of fouls and the false-call rate are identified too (Section~\ref{sec:4}). In the registered analysis this propensity rises with the calling score (logit slope 0.390 {[}0.220, 0.577{]}); the observed called share runs from about 0.6 in the lowest eighth of the score to about 0.9 in the highest (Figure~\ref{fig:2}), and part of the rise lies between foul types; within types it is suggested, not established (exploratory; Section~\ref{sec:5.4}). The programme does not guarantee completeness. We treat it as a baseline, give each conclusion the weakest premise about the listing it needs, and report the rest as bounds and tipping points \citep{manski2003}. \emph{Third, what a test inside the listed sample answers.} By its pre-registered criterion, a calling score from pre-whistle geometry and game context shows no strong directional separation of the listed missed fouls from the listed correct non-calls (area under the ROC curve, AUC, 0.469 {[}0.415, 0.526{]}). The result concerns ranking among the plays the reviewers chose to list, not whether the score carries information about fouling; under complete listing the data answer that question directly, under a proportional-selection scenario, in which missed fouls are listed in proportion to non-fouls at the same score, they cannot (Sections~\ref{sec:4.5} and~\ref{sec:5.5}). We recognised this distinction only after registration. The body follows the order of the registered plan, with the test (Section~\ref{sec:5.3}) before the propensity (Section~\ref{sec:5.4}).

We use the one season, 2015-16, in which, to our knowledge, public player tracking at 25 Hz overlaps the league's published assessments: 223 games, 2,317 contact-foul plays of the reports and 170 listed missed calls, of which 109 enter the registered tests as analysed records and 105 the propensity and the registered level as proximity segments. The tracking ends on 23 January 2016, so every estimate that joins the tracking to the reports describes the October--January reports; comparisons across months and seasons use the reports alone. Players are points on the court plane, without limbs or height, and the officials are not tracked, so every rule condition is a proxy. Beyond the three points, the study gives, to our knowledge, the first segment-level alignment of public tracking, play-by-play and the league's assessments, and catalogues eight mechanisms by which a review leaks into sensor-derived features, quality rules or sampling frames (Section~\ref{sec:7.3}; supplement, Section S4).

\emph{Registered and exploratory.} We registered the analysis plan on the Open Science Framework on 2026-10-01 (osf.io/sr243; Appendix~\ref{app:A}) and ran the registered analyses with the archived code; Appendix~\ref{app:A} records one narrowed reading of the main test and the sign convention of a transport parameter taken from the code. The plan was written after the pilot (supplement, Section S12) and before any relation between a tracking feature and a league assessment was computed on non-pilot plays; the registered document discloses what had been seen and what was adjusted on that basis (supplement, Section S7.1). A second plan, for the follow-up of Section~\ref{sec:5.7}, was registered the same day (osf.io/hnzdc), after the results of the first were known; it has one reporting deviation (Appendix~\ref{app:A}). Both registrations are under embargo until 2028-10-01; referees can read them through the anonymised view-only links above. Everything added after the results were known is labelled \emph{exploratory}. Table~\ref{tab:6} grades the principal empirical results: \emph{confirmatory}, a registered analysis with a registered decision rule; \emph{registered, descriptive}, a registered analysis reported without one; and \emph{exploratory}.

We make no claim about any player's intent and none about any official: officials' names are removed when the data are read, no official-level quantity is estimated, and the released tables add no official-level information to what is public.

\section{The review and the observation mechanism}\label{sec:2}

\subsection{The Last Two Minute reports}\label{sec:2.1}

The programme began in March 2015. For close games the league publishes an assessment of the officiating in the last two minutes of the fourth period and, in 2015-16, of each overtime period: games ``\ldots{} which were within five points at the two-minute mark (and during overtime, where applicable)'' \citep{nbal2m20160219}. Every game within 5 points at 2:00 of the fourth period or an overtime has a report (406); the other 4 of the 410 reports are games 6--10 points apart at 2:00 that came within 3 later, added at the league's discretion (supplement, Section S2.1). A report lists every call in that span and the non-calls the reviewers consider \emph{material}, ``material plays directly related to the outcome of a possession'' \citep{nbal2m20160223}, in practice the ``more obvious or noticeable plays, those deemed controversial or being talked about'' \citep{nbamemo2016}: a qualitative criterion of materiality and salience that does not fix the probability that a given non-call is listed. Each play receives a comment and one of four assessments, correct call (CC), incorrect call (IC), correct non-call (CNC) or incorrect non-call (INC), marked incorrect only on ``clear and conclusive'' video evidence \citep{nbal2mfaq}, by the league's senior officiating management after the game, with several camera angles and knowledge of the outcome. We call them the \emph{league assessment}: the most authoritative public verdict on an officiating decision, but a verdict rather than a measurement. Criterion, coverage and practice have changed since 2015-16 (supplement, Sections S2.1 and S14), and within 2015-16 itself (Section~\ref{sec:6}).

\subsection{Related work}\label{sec:2.2}

Economic studies of officiating have mostly used box-score foul counts and outcomes \citep{price2010,price2012,rodenberg2011}. The reports opened a play-level route: \citet{deutscher2015}, \citet{gong2022}, \citet{pelechrinis2023}, \citet{mocan2024}, \citet{belasen2025} and \citet{sabag2026} study bias, home advantage, betting lines and coverage areas, and \citet{tu2026} models the probability that a listed non-call is a missed call (supplement, Section S2.5). All take the play as the unit and the league's assessment as the standard of correctness; we make this premise explicit. Work on NBA optical tracking concerns possession value and defence \citep{cervone2016,franks2015}, with \citet{bunker2025} on the public 2015-16 release; DeepHoops \citep{sicilia2019} predicts fouls from trajectories and soccer fouls have been classified from multi-view video \citep{held2023}, but neither relates a call to a review of it. In baseball the tracking itself judges the umpire \citep{hunter2018,deshpande2017}; a foul cannot be judged from centroids, so the league's review is the only source of correctness.

The review observes decisions on all units and correctness on a subset that includes every positive decision and some negative ones. Its relatives, tabulated in the supplement (Section S1), are verification-bias methods for diagnostic tests \citep{begg1983,zhou1998,alonzo2005,pepe2003}, two-phase designs \citep{neyman1938,breslow1988}, positive--unlabeled learning (\citealp{elkan2008}; \citealp{bekker2019}; survey: \citealp{bekker2020}), selective labels \citep{lakkaraju2017,kleinberg2018} and trusted-subset corrections of noisy labels \citep{hendrycks2018}. Non-ignorable verification is handled by selection models with sensitivity parameters or parametric models of verification \citep{kosinski2003,rotnitzky2006,fluss2009,liu2010}, and verifying only the positives identifies relative but not absolute sensitivity \citep{pepe2001}. Here every positive decision is verified and, under completeness, every missed one is listed, so the propensity is identified without a model for the verification of negatives, as in the empirical sensitivity of cancer screening \citep{day1985}; the selection of negatives enters through completeness and tipping points, in the spirit of \citet{rotnitzky2006}.

\subsection{Data, linkage and samples}\label{sec:2.3}

\textbf{Sources.} The 2015-16 regular season is, to our knowledge, the only NBA season whose optical tracking became public \citep{linouk232016}: 636 archived games played between 2015-10-27 and 2016-01-23, of which 631 parse, with the court-plane coordinates of the ten players and the ball at 25 Hz; the officials' positions are not public. The play-by-play \citep{sumitrodatta2022} records every foul with its sub-type and players. The assessments come from the community table of the reports in the repository \texttt{atlhawksfanatic/\allowbreak{}L2M} \citep{atlhawksfanatic2025} at commit \texttt{330ab64a} (MIT License; Appendix~\ref{app:B}), checked by hand and by the match rate of called fouls. The rule predicates follow the Official Rules of the National Basketball Association, 2015--2016 edition \citep{nba2015rules}, quoted by rule, section and page. Officials' names are stripped when a file is read.

\textbf{Linkage.} \emph{Called plays} (CC, IC) are matched to a play-by-play foul in the same game and period, within 2 s of the report's clock, with the same fouling player; 99.1~\% of the season's 1,915 called fouls match. \emph{Non-called plays} (CNC, INC) are located in the tracking by searching the running-clock frames within \([t - 1\ \text{s}, t + 2\ \text{s}]\) of the reported clock \(t\) for the two named players and anchoring on their closest approach. Among usable rows, 88.7~\% of CC, 87.3~\% of CNC, 82.1~\% of INC and 95~\% of IC plays link to a proximity segment (supplement, Sections S2 and S10.2).

\textbf{Samples.} Table~\ref{tab:1} follows the season's reports down to the registered analyses. The 223 \emph{intersection games} have a report and a non-empty tracking archive and hold 2,317 contact-foul rows once the non-contact fouls, eight other fouls and the rows the report text types as take fouls are removed. Take fouls are a category of both the reports and the play-by-play; they arise in \emph{end-game fouling situations}, which we define by the score, the clock and the team-foul state, never by an inference of intent (supplement, Section S9.3). Of these, 1,998 are \emph{usable} (at least half of the 2 s before the play tracked); 1,704 of these link to a proximity segment and are not among the 50 \emph{pilot} plays examined before registration (supplement, Section S12); setting aside 12 that share their segment with a take foul leaves 1,692 \emph{analysed records}.

\begin{table}[!htbp]
\centering
\caption{From the season's reports to the analysed records.}
\label{tab:1}
\small
\begin{tabular}{>{\raggedright\arraybackslash}p{0.60\linewidth} >{\raggedright\arraybackslash}p{0.34\linewidth}}
\toprule
Step & Count \\
\midrule
Games with a public tracking archive & 636 (631 parse; 4 empty, 1 unreadable) \\
Games with a report, 2015-16 regular season & 410 \\
\ldots{} in the tracking date range & 231 \\
\ldots{} with a tracking archive & 225 (6 missing) \\
\ldots{} with a non-empty archive (intersection) & 223 \\
Assessed plays in the 223 games (non-blank assessment) & 3,153 (3,050) \\
\ldots{} foul rows & 2,738 \\
\ldots{} after removing take fouls (381), non-contact fouls (32) and other fouls (8) & 2,317 contact-foul rows (2,283 assessed: CC 591, IC 47, INC 170, CNC 1,475) \\
\ldots{} with $\ge$ 0.5 pre-play tracking coverage (usable) & 1,998 \\
\ldots{} linked to a proximity segment, pilot plays excluded (confirmatory sample) & 1,704: INC 109 / CNC 1,104 / CC 453 / IC 38 \\
\ldots{} after setting aside the 12 plays linked to a take foul (analysed records) & 1,692: INC 109 / CNC 1,101 / CC 445 / IC 37 \\
Sensitivity sample (usable, linked or not, pilot excluded) & 1,948 \\
\bottomrule
\end{tabular}
\end{table}

\subsection{Variables and units}\label{sec:2.4}

\begin{figure}[!htbp]
\centering
\includegraphics[width=0.78\linewidth]{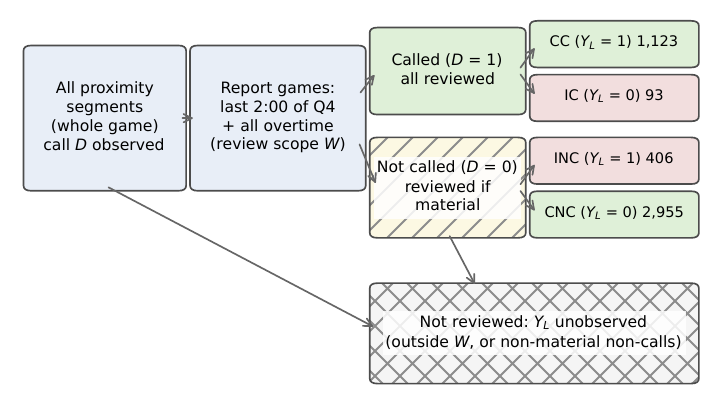}
\caption{The review structure. Every contact in a game is either called (D = 1) or not (D = 0), and D is observed for all of them. Inside the review scope W every call is assessed as correct (CC, \(Y_L = 1\)) or incorrect (IC, \(Y_L = 0\)); of the non-calls, only those the reviewers judge material are assessed, as missed calls (INC, \(Y_L = 1\)) or correct non-calls (CNC, \(Y_L = 0\)). Hatched: observed only through the reviewers' selection; cross-hatched: unobserved (outside W, or non-calls the reviewers do not list). Counts: the 2015-16 regular season's assessed contact fouls in the 410 report games (take fouls and unassessed plays removed); in the 223 games with tracking: 591 / 47 / 170 / 1,475 (Table~\ref{tab:1}). The calling propensity \(e(x) = P(D = 1 \mid Y_L = 1, x)\) is identified from CC and INC alone if every missed foul in W is listed.}
\label{fig:1}
\end{figure}

Table~\ref{tab:2} states the variables and units. \(D\) records whether a contact was called and \(Y_L\) the league's assessment, whether the league would grade the contact a foul had it reviewed it; the assessment categories are the cells of \((D, Y_L)\): CC \((1,1)\), IC \((1,0)\), CNC \((0,0)\), INC \((0,1)\). \(W\) is the \emph{review scope}: in a game with a report, the last two minutes of the fourth period and every overtime period. A \emph{true foul} has \(Y_L = 1\); a \emph{missed call} is an INC play, a \emph{missed foul} any uncalled true foul, listed or not. Figure~\ref{fig:1} shows the structure: entry into the scope is a known function of the game, every call inside it is listed, and non-calls are listed by an unrecorded judgement of materiality.

\begin{table}[!htbp]
\centering
\caption{Variables, the quantities built from them, and the three units of count, with what is observed or identified. \(x\) is the pre-whistle geometry and context of a segment (Section~\ref{sec:5.1}). The estimands with their reporting devices are in the supplement (Table S3).}
\label{tab:2}
\small
\begin{tabular}{>{\raggedright\arraybackslash}p{0.22\linewidth} >{\raggedright\arraybackslash}p{0.40\linewidth} >{\raggedright\arraybackslash}p{0.27\linewidth}}
\toprule
Symbol or unit & Meaning & Observed or identified \\
\midrule
\(D\) & Called on the floor & Observed on every segment \\
\(Y_L\) & The league's assessment: a foul as the league grades it & On listed plays; on unlisted segments only through completeness (Section~\ref{sec:4.1}) \\
\(Y^{*}\) & A foul by the rule & Never (Section~\ref{sec:4.6}) \\
\(V\) & Listed in a report & Observed; every call inside \(W\) is listed \\
\(R\) & Retained by the project: usable, linked, not a pilot play & Observed, by assessment (Section~\ref{sec:4.2}) \\
\(L\) & A listed play linked to a detected segment & Observed for listed plays (Section~\ref{sec:2.3}) \\
\(W\) & The review scope (below) & Known for every segment \\
\(z\) & The calling score, \(\operatorname{logit}\hat p(x)\) & Out of fold on every segment (Section~\ref{sec:5.1}) \\
Calling probability \(p(x)\) & \(P(D = 1 \mid x)\) & Yes; cross-fitted model on all proximity segments (Section~\ref{sec:5.1}) \\
Calling propensity \(e(x)\) & \(P(D = 1 \mid Y_L = 1, x)\), inside \(W\) & The counts bound the play-level share \(\eta_W^{\text{play}}\) (Section~\ref{sec:3}); \(e(z)\) is estimated under the assumptions of Section~\ref{sec:4} \\
Prevalence \(\pi(x)\) and false-call rate \(f(x)\) & \(P(Y_L = 1 \mid x)\) and \(P(D = 1 \mid Y_L = 0, x)\), inside \(W\) & Under completeness (Section~\ref{sec:4.2}) \\
Completeness \(\kappa(x)\) and transport \(\delta\) & \(P(V = 1 \mid D = 0, Y_L = 1, x, W)\); \(\operatorname{logit} e_{\text{in}} - \operatorname{logit} e_{\text{out}}\) & No: grids and tipping points (Section~\ref{sec:4.3}) \\
\textbf{Unit: play} & A row of a report; in the review scope of the 223 games, 591 CC, 47 IC, 170 INC and 1,475 CNC & The count bound (Section~\ref{sec:3}); the registered tests count analysed records, plays retained by the project \\
\textbf{Unit: detected segment} & A proximity segment of an offence--defence pair; 24,513 in the review scope, 24,393 after removing the 120 that the reports list as take or other fouls, of which 508 are called & The prevalence and the listing models; the registered propensity fits on the 24,513 plus nine (Sections~\ref{sec:4} and~\ref{sec:5}) \\
\textbf{Unit: augmented segment} & 24,393 matched segments plus each call or miss not linked to a matched segment (124 CC, 9 IC, 45 INC) as one: 24,571; no CNC added & The segment-level levels (supplement, Table S45) \\
\bottomrule
\end{tabular}
\end{table}

\section{A bound from the report counts}\label{sec:3}

Of all true-foul plays in the review scope of the 223 intersection games (October--January), listed or not, the report counts put the called share at about 78~\% at most. The bound refers to the contact-foul plays of the reports as filtered here, not to NBA fouls in general; it was computed after registration and is exploratory.

\emph{The count.} In the intersection games the reports list 591 correct calls, 47 incorrect calls and 170 missed calls among the assessed contact fouls of the review scope. Of the 34 contact-foul plays listed without an assessment, 3 have a play-by-play foul by the named player within 3 s; for two of them that foul is also the foul of a correct call listed for the same player at (nearly) the same clock, so the play-by-play check below counts only one as an unassessed call (its second unassessed row is a take foul). The called share of listed true-foul plays is 591/(591 + 170) = 0.777 {[}0.744, 0.806{]} (2,000 game-cluster resamples). It is exact for the listed plays and, under complete listing, for all true-foul plays. The interval is the sampling interval of the bound statistic, not a two-sided identified interval; as an estimate for games like these the share is at most 0.806, a one-sided 97.5~\% limit. The bound constrains \(\eta_W^{\text{play}}\), the calling propensity averaged over the true-foul plays of the scope, not the propensity \(e_W(z)\) at each calling score.

\emph{What it needs, and what it does not.} If every missed foul in the review scope is listed (A1), the share of true fouls that were called is 0.777 {[}0.744, 0.806{]}, and the listed 0.76 {[}0.64, 0.90{]} missed calls per game are all the missed true fouls; if listing is incomplete, the share is lower and the missed fouls more numerous: over a completeness of 0.6--1, 0.68--0.78 and 0.76--1.27 per game (supplement, Table S43). In general, if a fraction \(\kappa\) of the uncalled true fouls is listed, the share called is CC/(CC + INC/\(\kappa\)): 0.736, 0.676, 0.582 and 0.511 at completeness 0.8, 0.6, 0.4 and 0.3 (supplement, Section S10.1). Since completeness cannot exceed 1, unlisted missed fouls can only enlarge the denominator, so 0.777 is an upper bound under any listing. It needs only that every call inside the scope is listed, as the programme states and the data confirm: every called review-scope segment without any linked record corresponds to a listed play (supplement, Section S10.1), and, checked from the play-by-play side, every one of the 1,021 non-technical fouls in the review scope of the 223 games appears in the reports (1,017 as a call of the same player; 2 as unassessed rows; 2 attributed to another player at the same clock); and that take fouls are removed as implemented (A4, Section~\ref{sec:4.1}); it needs no tracking, linkage, model or complete listing. It is the upper general bound for sensitivity of \citet{kosinski2003b} when every positive test is verified, whose test ignorance region then reduces to a curve indexed by our completeness \(\kappa\), and a worst-case bound in the sense of \citet{manski2003}, not a new result. Counting the 3 unassessed called plays as correct calls leaves 0.777 {[}0.745, 0.807{]}; even if all 34 unassessed plays had been correct calls, the share would be 0.786.

\emph{No lower bound of practical use.} With every uncalled review-scope segment a true foul, the formal lower bound is 0.024 {[}0.022, 0.026{]} on the augmented segments (Table~\ref{tab:2}; supplement, Table S45). Under the proportional-selection scenario of Section~\ref{sec:5.5} the share runs from 0.14 to 0.58, depending on the unidentified scale of completeness (supplement, Table S31). Uncalled true fouls per game are bounded only from \emph{below}, by the 0.76 {[}0.64, 0.90{]} listed missed calls per game.

\emph{Other ways of counting.} Over detected segments, a model on the linked segments gives 0.784 {[}0.745, 0.820{]} (Section~\ref{sec:4.2}), and the registered segment model 0.791 {[}0.749, 0.825{]}, kept as the registered record. Both need a model and are computed on detected segments, not on the listed plays, and both lie above 0.777: the linked-segment level for reasons of population, the registered 0.791 also because missed calls are retained less often than correct calls among the analysed records (Section~\ref{sec:4.2}; supplement, Section S10.1). The play-level 0.777 needs neither linkage nor a model and is the level this paper reports. Variants by population, label and weights are in the supplement (Section S10.1).

\section{What completeness identifies}\label{sec:4}

\emph{The idea in one paragraph.} Every call inside the review scope is listed. If every uncalled true foul is listed as well (completeness), the listed true fouls are all the true fouls, and the share called and its variation with the pre-whistle geometry follow without the listed correct non-calls, whose selection is the unrecorded part of the review. The rest of this section says what is lost when completeness fails.

\subsection{Assumptions, and how each conclusion uses them}\label{sec:4.1}

The registered plan states five assumptions, each with a parameter or a check.

\begin{itemize}

\item
  \textbf{A1, completeness.} Every uncalled true foul inside \(W\) is listed as an incorrect non-call: \(\kappa(x) = P(V = 1 \mid D = 0, Y_L = 1, x, W) = 1\); registered grid \(\{0.6, \ldots, 1\}\).
\item
  \textbf{A2, no false calls.} \(f = 0\); registered grid \(\{0, 0.01, \ldots, 0.05\}\). A2 is not needed: since every call is listed, A1 identifies the false-call rate, about 0.002 per non-foul segment (Section~\ref{sec:4.2}).
\item
  \textbf{A3, transport.} The propensity inside \(W\) equals that outside it; \(\delta\) on the logit scale, \(\{-0.5, 0, 0.5\}\). The main results are stated for \(W\) only.
\item
  \textbf{A4, take fouls removed.} Three rules apply: report rows typed as take fouls by the report text are removed from the play counts (Table~\ref{tab:1}); segments whose play-by-play foul carries the take-foul tag are removed from every segment population; and the 12 analysed records linked to such a segment are set aside. The tag does not cover every end-game fouling situation: inside the review scope, 117 called segments that the reports type as take fouls carry no such tag (supplement, Section S10.1).
\item
  \textbf{A5, detection and linkage do not depend on the assessment given \(x\).} Retention without conditioning on \(x\) does differ by assessment (Section~\ref{sec:4.2}), so we report weighted sensitivity analyses and state the condition on relative retention each conclusion needs.
\end{itemize}

We treat A1 as a \emph{baseline}, not a belief: under it the propensity, the prevalence and the false-call rate are identified, and we measure every departure from it. It may fail because the materiality criterion applies to non-calls whether or not an infraction occurred (Section~\ref{sec:2.1}): a missed foul judged immaterial is not listed. Since listing depends on how material and how noticeable a non-call is, completeness may vary with the play, \(\kappa(x)\), in a form the league's materials do not fix; it may also differ by foul type (Section~\ref{sec:5.6}) and within the season (Section~\ref{sec:6}). If only a fraction \(\kappa(x)\) of missed fouls is listed, the regression of \(D\) on \(x\) among CC and INC estimates

\[
\tilde e(x) \;=\; \frac{e(x)}{e(x) + \kappa(x)\,\bigl(1 - e(x)\bigr)} \;\ge\; e(x),
\]

so incompleteness inflates the estimated propensity. We do not resolve whether A1 holds; we state, for each kind of conclusion, the weakest premise about the listing it needs.

\begin{itemize}

\item
  \emph{Levels.} The play-level count is an upper bound under any listing (Section~\ref{sec:3}). The model-based levels also need the single-index propensity of Section~\ref{sec:4.3} with completeness constant or log-linear in the score, A5, and true fouls that appear as proximity segments. The level on linked segments is weighted for retention within the detected frame (Section~\ref{sec:4.2}); the registered 0.791 is unweighted, and the other variants are in the supplement (Section S10.1).
\item
  \emph{The direction of the slope} needs only A5 and a completeness that does not fall with the calling score: a missed foul that looks like a usually called contact must not be \emph{harder} to list. Since \(\operatorname{logit}\tilde e(z) = \operatorname{logit} e(z) - \log\kappa(z)\), a constant \(\kappa\) shifts the level only, and a \(\kappa\) rising with the score lowers (and can reverse) the observed slope; the observed slope is then a lower bound on the true one, exactly so under the linear-logit forms of Section~\ref{sec:4.3}. A completeness that differs by foul type, where types differ in score, is one way it can fall with the score, and the log-linear form of Section~\ref{sec:4.3} does not cover it.
\item
  \emph{Differences between groups} need no A1, only the \emph{ratio} of completeness between the groups and equal relative retention of correct and missed calls in them; the score-adjusted ratios also need a completeness gradient in the score that is common to the groups or log-linear (supplement, Section S1.3).
\end{itemize}

Under proportional selection (Sections 4.7, 5.5) the propensity is not point identified and the constant-completeness grid and the envelopes do not apply; the count bound of 0.777 still holds, and a positive observed slope is still a lower bound as long as completeness does not fall with the score.

\subsection{What completeness identifies, and what the data add}\label{sec:4.2}

Every call inside \(W\) is listed, \(P(V = 1 \mid D = 1, W) = 1\), and under A1 every uncalled true foul is too. The union of CC and INC is then \emph{all} the true fouls in \(W\), and

\[
e(x) \;=\; P(D = 1 \mid Y_L = 1, x, W)
\]

is estimated by regressing \(D\) on \(x\) in the CC and INC rows. No correct non-call enters, so the selection function \(q(x) = P(V = 1 \mid D = 0, Y_L = 0, x, W)\) affects only the composition of the CNC set and cannot bias \(e(x)\). Since an unlisted non-call then has \(Y_L = 0\), every segment in \(W\) is classified, and the prevalence and the false-call rate are identified as well. With \(p(x) = e(x)\,\pi(x) + f(x)\,(1 - \pi(x))\),

\[
\pi(x) \;=\; \frac{p(x) - f(x)}{e(x) - f(x)},
\]

with \(p\) recalibrated on the review-scope units, and \(\pi(x) = p(x)/e(x)\) at \(f = 0\).

\emph{False calls.} Because every call is listed and true fouls are a few per cent of the segments, the false-call rate hardly moves with \(\kappa\): the 47 incorrect calls (7.4~\% of the 638 calls) give 0.0020--0.0021 per non-foul segment for \(\kappa \ge 0.1\). The registered grid of \(f\), 0.01--0.05, is incompatible with the 47 incorrect calls for any completeness of at least 0.01 (a post-hoc observation). Over the same range of completeness, the share called on the population of the count bound is 0.68--0.78 and the missed true fouls per game 0.76--1.27 (Section~\ref{sec:3}). The registered computation on the 24,522 bounds units, which include the 120 called segments listed as take or other fouls, gives other values (Appendix~\ref{app:A}); segment-level per-game figures, such as Table~\ref{tab:4}'s 0.61 under complete listing, count only the true fouls captured as segments in the 222 detected games, and so lie below the 0.76 listed per game. The registered envelope is archived in Appendix~\ref{app:A} (supplement, Table S47). Among the calls, incorrect calls are not negligible, and their share along the score is the check on \(f = 0\) for the prevalence gradient (supplement, Section S8.4 and Table S34). Because some called segments carry no record and some listed plays link to no segment, the segment-level quantities also rest on the linkage rule and an assumption about unlinked plays (supplement, Section S10.1); the play-level count does not.

\emph{The level on linked segments.} On the 24,393 review-scope segments matched to the report plays (all review-scope segments except those linked to the 120 plays the reports list as take or other fouls), with the play-by-play call label and the propensity weighted for retention within this detected frame (the pilot draw, unusable windows and failed links inside the frame count as losses, plays outside the frame do not), the model gives 0.784 {[}0.745, 0.820{]} of true fouls called (the interval holds the retention weights fixed; re-estimating them changes it by less than 0.005) and 0.63 {[}0.50, 0.77{]} missed per game, the latter among the detected segments of 222 games (supplement, Table S45). It lies above 0.777 because it is computed on the plays the detector captured in the frame, which leave out more of the listed missed calls (42 of 170 lie outside the detected frame) than of the correct calls (124 of 591); the supplement reconciles the two step by step (Table S45, Panel D). Weights taken over all report rows give 0.779 {[}0.739, 0.815{]} instead; they correct the propensity with losses from outside the frame, and so hold only if the out-of-frame plays can be transported into the frame (supplement, Sections S8.4 and S10.1). The registered 0.791 is the same computation on the 24,522 segments of the bounds analysis (the 24,513 review-scope segments and 9 linked segments just outside by their own clock), unweighted, and is kept as the registered record. Both segment populations differ from the population of listed plays behind the count bound 0.777, which needs neither linkage nor a model and is the level we report.

\emph{Retention.} Comparing the report rows with the analysed records of Table~\ref{tab:1}, the intersection games retain 0.753 of correct calls, 0.787 of incorrect calls, 0.641 of missed calls and 0.746 of correct non-calls; missed calls are retained less often because of the pilot draw and lower linkage. Weighting by assessment corrects the overall difference but not the slope: with retention \(r_{\text{CC}}(z)\) and \(r_{\text{INC}}(z)\) along the score,

\[
\operatorname{logit} e_{\text{obs}}(z) \;=\; \operatorname{logit} e(z) - \log\kappa(z) + \log\frac{r_{\text{CC}}(z)}{r_{\text{INC}}(z)},
\]

so the slope's direction also needs a retention ratio that does not rise with the score (a falling ratio only understates the slope), and a group difference needs equal relative retention. Only usability can be checked against the score, and it is uninformative for missed calls, since a linkage failure has no score (supplement, Section S10.2).

\emph{Linkage.} A listed play enters the segment analyses only if it links to a detected segment, so ``listed'' in the segment models means \(\tilde V = V L\). It describes the league's listing only if the ratio of the linkage probabilities of missed calls and correct non-calls is constant in the score. Constant linkage of correct and missed calls adds only an intercept to the logit propensity (slope unchanged), and the ratio for missed calls and correct non-calls rescales \(c\) (Section~\ref{sec:5.5}). The intercept does not rescale \(\pi(z) = p(z)/e(z)\) uniformly, so the prevalence has no general guarantee. In these data the league readings agree with the joint ones (supplement, Table S32); a ratio that varies with the score would need an assumption of its own. Unlinked rows are classified in Section S10.2; Section~\ref{sec:6} reports what the data say about the condition.

\subsection{Bounds and tipping points: the specification}\label{sec:4.3}

We do not estimate completeness or transport; the registered plan fixes the following (units, recalibration and bootstrap in the supplement, Section S7.1). The propensity is a function of the calling score alone, fitted on the listed true fouls with INC weighted by \(1/\kappa\):

\[
z = \operatorname{logit}\hat p(x), \qquad \operatorname{logit} e(z) = a + b z, \qquad \pi(z) = \frac{P_W(D \mid z) - f}{e(z) - f} \in [0, 1],
\]

with \(P_W\) the recalibration of \(\hat p\) on the review-scope units. A uniform under-listing changes every group's CC-to-INC odds by the same factor, so it cannot reverse a difference between groups. Only completeness that differs between a configuration and the rest can, and the difference vanishes at

\[
\rho^{*} \;=\; \frac{\kappa_{\text{in}}}{\kappa_{\text{out}}} \;=\; \frac{\text{CC}_{\text{out}}/\text{INC}_{\text{out}}}{\text{CC}_{\text{in}}/\text{INC}_{\text{in}}}, \qquad \rho^{*}_{\text{adj}} = \exp(-\theta)\ \text{ for } \theta \text{ in } \operatorname{logit} e = a + bz + \theta\,\mathbf{1}\{\text{inside}\},
\]

the registered raw form and the score-adjusted form. We read each against the range a completeness grid can produce: 0.60--1.67 under the registered grid and 0.30--3.33 with the grid extended to \(\kappa = 0.3\). These ranges are reporting devices, not estimates; Section~\ref{sec:5.4} sets them against the within-season evidence. For the slope \(b\) the relevant departure is \(\kappa(z) \propto \exp(\gamma z)\): the complete-data slope is then \(b + \gamma\) at any scale and the cancelling gradient is \(\gamma^{*} = -b\), exact for the population logistic model with these linear-logit forms. Both linear-logit fits agree well with the observed shares overall, with a local departure that the direct counts show too (supplement, Section S10.3). Scale-dependent quantities need \(\kappa(z) = \min\{1, \kappa_0 \exp[\gamma (z - z_0)]\}\) (Section S8.5). We report levels as counts, as model-based values at \((f, \kappa) = (0, 1)\) with a game-clustered interval, and as envelopes over the grids, in the sense of partial identification \citep{manski2003}.

\subsection{Comparisons built from the reports}\label{sec:4.4}

Any study that uses the reports' assessments to compare groups (home and away, seasons, foul types) inherits the unrecorded materiality judgement whenever a listed correct non-call enters its numerator or denominator. A comparison built from calls alone or from true fouls alone is free of the selection of \emph{non-fouls} (supplement, Table S4) and depends on completeness only through its ratio between the groups. With \(u_g\) the share of true fouls among a group's uncalled contacts in \(W\) and \(q_g\) the probability that its correct non-call is listed, \(\text{INC}_g/\text{CNC}_g = (\kappa_g/q_g)\,u_g/(1 - u_g)\): a comparison of these odds needs the same \(\kappa_g/q_g\) in each group, and reading the share as a miss rate needs \(\kappa_g = q_g\). \citet{tu2026} estimates, by category of play, the share of listed non-calls that were missed calls; compared across categories, that share needs the same \(\kappa_g/q_g\) in each, which the reports do not record. This concerns the assumptions such comparisons need, not any study's conclusions.

\subsection{The main test, and what it answers}\label{sec:4.5}

The registered plan first asked whether the pre-whistle calling score separates the listed non-calls that were missed fouls from those that were correct: the AUC (area under the ROC curve) of \(\hat p(x)\) for INC against CNC among the analysed records. Two hypotheses must be kept apart. (a) Among the listed plays, missed fouls and correct non-calls cannot be told apart on the score; then the AUC is one half with no further condition. (b) The score carries no information about fouling among all segments. Under selective listing (b) does not imply an AUC of one half, for two reasons: (i) if the reviewers list non-fouls that look like calls and every missed foul, the listed correct non-calls look more like calls than the listed missed fouls; (ii) under any listing, the test compares non-calls only, and when the propensity rises with the score the uncalled true fouls concentrate where the score is low. The registered rule concerns (a): informative if the 95~\% interval lies above 0.65 or below 0.35, insufficient if it lies inside (0.35, 0.65), inconclusive otherwise. A value near one half says the score shows no strong directional separation among the listed plays; it cannot answer (b). A test of (b) needs benchmarks in which \(e\) and \(q\) are fixed in advance, estimated on independent units, or estimated jointly under an inference that refits every step; under completeness it reduces to a test of the prevalence gradient, which the data identify directly (Section~\ref{sec:5.5}). Every statistic that uses the correct non-calls (the main test, SA1, SA3, SA4, and the review budget, which asks whether the score flags the assessments that were errors) is conditional on listing.

\subsection{Grading error: the league's assessment and the rule}\label{sec:4.6}

Every estimand above is defined with \(Y_L\). Whether the assessments are right by the rule, \(Y^{*}\), has three parts the data do not identify. A listed missed call may not be a foul by the rule, which understates the share called with respect to \(Y^{*}\). A call graded correct may be a false call by the rule, since overturning a call needs clear and conclusive evidence; we carry this as an under-listing factor \(m \in \{1, 1.5, 2\}\) for incorrect calls, a judgement introduced after registration, which lowers the share called by about 0.01--0.02 and leaves the missed fouls per game essentially unchanged (within 0.001) (supplement, Table S49). A foul by the rule on inconclusive video is graded a correct non-call; this gap between \(Y_L\) and \(Y^{*}\) is not a failure of completeness and may put the share called with respect to \(Y^{*}\) below the bound of Section~\ref{sec:3} by an amount the data do not bound (Table S46, Panel B; Section S1.5). None of the three touches our conclusions about the league's assessments.

\subsection{Checks by simulation}\label{sec:4.7}

Two simulations check the argument (supplement, Section S5). In the first, with known truth and the review structure of Section~\ref{sec:2.4}, treating non-calls as non-fouls understates the prevalence of fouls by 0.07 to 0.10 in every setting, while \emph{given completeness} the review-propensity estimator is unaffected by the selection of non-fouls; the bounds contain the true value for 96--99~\% of units (a share of units, not a confidence coverage) but are wide, so tipping points are the primary sensitivity summary. The second, exploratory and calibrated to the review scope as in \citet{brill2025}, shows that one half is not a universal value for an uninformative score: it is 0.17 with complete listing of missed fouls and non-fouls selected by materiality. Under proportional selection, at the base slope the estimated slope changes sign (\(-0.417\) against 0.39) and the share called is 0.977 against a truth of 0.843; when completeness rises with the score, the slope is biased toward zero by 0.107.

\section{The conditional analysis on the tracking data}\label{sec:5}

\subsection{Segments and the calling score}\label{sec:5.1}

A foul involves two players, so the unit of analysis is a \emph{proximity segment} of an offence--defence pair: on running-clock frames with ten players on the court, a maximal run of consecutive frames with the pair's centroid distance at most 4.0 ft, containing at least two frames at 3.0 ft or less. Most segments are ordinary physical traffic, not contact in the rulebook's sense. A segment is \emph{called} (\(D = 1\)) if a play-by-play contact foul by or on the pair falls within 2.5 s before its start to 0.5 s after its end (supplement, Section S3.1). Over the 631 games the detector yields 1,052,162 segments and recovers 86.8~\% of called fouls; 657 lie at take fouls, and 21,358 of the remaining 1,051,505 (about 2~\%) are called. Every feature is computed on the second before the earlier of the closest approach and the estimated whistle onset, and a code-base test fails if any feature reads a later frame. The features describe geometry (pair, defender, ball, screener) and context (game and shot clock, score margin, team-foul state, home court); the rule predicates (legal guarding position, restricted area, upward motion, and an off-path proxy for a legal screen) are proxies computed on the same frames (supplement, Sections S3.3--S3.4 and Table S8).

We estimate the calling probability \(p(x) = P(D = 1 \mid x)\) by \(L_2\)-regularised logistic regression with the play-by-play calls as labels, pooling foul types by necessity, since uncalled segments have none. Predictions are cross-fitted by game in five folds, so every calling probability is out of fold; the later fits on the review scope (\(P_W\), \(e\), the listing model) are in sample. The \emph{calling score} \(z = \operatorname{logit}\hat p(x)\) is the single index of Section~\ref{sec:4.3}; no league assessment enters it. A gradient-boosting model with the same features is the score of SA5. Geometry carries essentially all the discrimination (Table~\ref{tab:3}). The full model separates called from uncalled segments at 0.793 inside the review scope and at 0.622 among contested on-ball drives, so most of the headline 0.841 separates contact situations from incidental proximity. The model describes where officials call, not where fouls occur (supplement, Section S6).

\begin{table}[!htbp]
\centering
\caption{Out-of-sample discrimination of called from uncalled proximity segments (ROC-AUC), by model, feature set and population. Registered were the specifications of the main logistic model and of the gradient-boosting model. Before registration, the out-of-fold discrimination of the calling model on play-by-play labels was known and disclosed in the registration: 0.841 on all segments and 0.793 inside the review scope (gradient boosting 0.866 and 0.854); it had never been evaluated against an assessment. The AUC values are descriptive; the geometry-only and context-only rows and the contested-drive column are exploratory. Labels are play-by-play calls. Contested on-ball drives: segments in which the offensive player has the ball and moves toward the basket at more than 5 ft/s while the defender is within 8 ft of the basket. Review-scope call share: 2.6~\% (628 segments) by the play-by-play label, 2.5~\% (617) with the report-record label of the level computations.}
\label{tab:3}
\small
\begin{tabular}{>{\raggedright\arraybackslash}p{0.28\linewidth} >{\raggedright\arraybackslash}p{0.17\linewidth} >{\raggedright\arraybackslash}p{0.17\linewidth} >{\raggedright\arraybackslash}p{0.21\linewidth}}
\toprule
Model, feature set & All segments (1,051,505; 2.0~\% called) & Review scope (24,513; 2.6~\% called) & Contested on-ball drives (22,022; 15.3~\% called) \\
\midrule
Logistic regression (main), full & 0.841 & 0.793 & 0.622 \\
Logistic regression, geometry only & 0.841 & 0.794 & 0.619 \\
Logistic regression, context only & 0.543 & 0.530 & 0.491 \\
Gradient boosting (SA5 score), full & 0.866 & 0.854 & 0.674 \\
\bottomrule
\end{tabular}
\end{table}

\subsection{Sample and inference}\label{sec:5.2}

The registered sample is the 1,692 analysed records of Section~\ref{sec:2.3}: 445 CC, 37 IC, 1,101 CNC and 109 INC. The registered secondary analyses are SA1, the main test on shooting fouls and screens; SA2, CC against INC among listed true fouls; SA3, the score against the share of listed plays that should have been called; SA4, the enrichment of INC in three risk configurations (Section~\ref{sec:5.6}); and SA5, the main test with the gradient-boosting score. Four negative controls N1--N4, contrasts for which the plan registered an expected ordering or no separation (supplement, Table S23), and a home-court example complete the plan. Intervals are game-cluster bootstrap intervals that refit the review-scope recalibration \(P_W(D \mid z)\) and the propensity \(e(z)\) in each resample, with the calling score held fixed. Refitting the first-stage model too (B = 200) leaves the intervals of the slope and the cancelling ratio essentially unchanged (standard-deviation ratio 0.97--0.99); the main test's AUC and the configuration contrasts are conditional on the first-stage score.

\subsection{The main test}\label{sec:5.3}

\emph{Plan.} Pooling all contact fouls, we compute the AUC of the calling probability for INC (positive class) against CNC, with a 95~\% percentile interval from 2,000 game-level bootstrap replications, and apply the three-zone rule of Section~\ref{sec:4.5} with no registered directional expectation. The registered power calculation \citep{hanley1982} gave the ``informative'' verdict a probability of about 0.41 at a true AUC of 0.70 and about 0.95 at 0.75.

\emph{Result.} The AUC is 0.469 {[}0.415, 0.526{]} on 109 INC and 1,101 CNC records in 216 games, inside (0.35, 0.65), the zone the registered rule labels \emph{insufficient}. The calling score shows no strong directional separation of the listed missed fouls from the listed correct non-calls. The statement concerns ranking among the listed plays by two calling scores trained to predict calls (Section~\ref{sec:4.5}), and does not exclude information that is not monotone in the score. SA5 gives 0.471 {[}0.411, 0.534{]}, and the thirteen registered sensitivity variants with an interval stay in the insufficient zone, between 0.443 and 0.484. Of the other registered AUC-type analyses only SA2 and N1 (SA2's contrast within two small strata) leave the neighbourhood of one half: among the listed true fouls the score separates the called from the missed at 0.640 {[}0.580, 0.699{]}, so the propensity is not flat in the score (Section~\ref{sec:5.4}).

\subsection{The calling propensity: slope and direction}\label{sec:5.4}

\begin{figure}[!htbp]
\centering
\includegraphics[width=\linewidth]{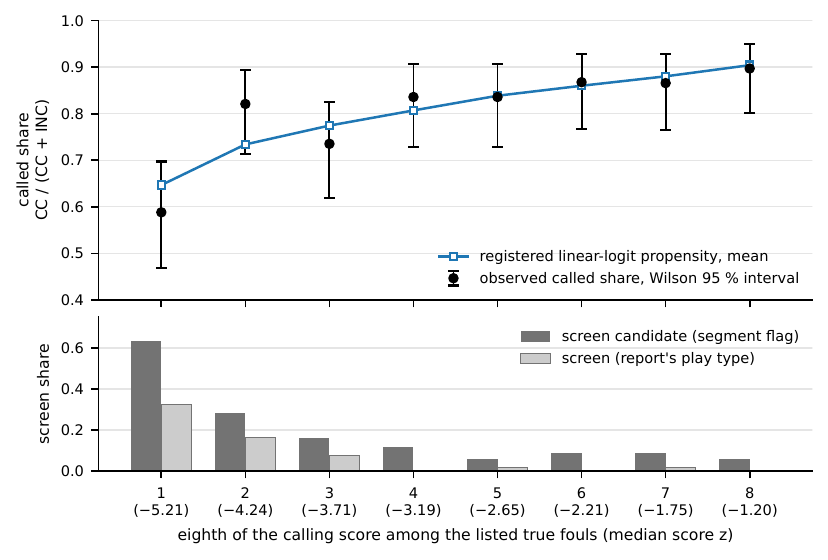}
\caption{The calling propensity along the calling score, and the foul types behind it. The 540 listed true-foul segments of the review scope (435 called, 105 missed), in eighths of the out-of-fold calling score z (67 or 68 each; median z in parentheses). Top: the observed called share CC/(CC + INC) with Wilson 95~\% intervals (points), and the mean of the registered single-index propensity, logit e = a + bz, in each eighth (line). Bottom (exploratory): the share of screens in each eighth, by the report's play type (from the comment text) and by the screen-candidate flag of the segment. The lowest eighth, where the called share is lowest, holds the largest share of the screens by either definition; part of the slope is a difference between foul types (Section~\ref{sec:5.4}; supplement, Section S10.5).}
\label{fig:2}
\end{figure}

A true foul that looks, to the on-floor model, like a usually called contact is itself more likely to have been called, under the conditions of Section~\ref{sec:4.1}. We run the specification of Section~\ref{sec:4.3} on 24,522 segments of 222 games (the 24,513 review-scope segments plus nine segments linked to a confirmatory record just outside the scope by their own clock); 1,660 are listed and 540 are listed true fouls. The slope of \(\operatorname{logit} e\) in \(z\) is 0.390 {[}0.220, 0.577{]} (registered computation, 500 resamples; {[}0.234, 0.561{]} with 2,000). Part of it lies between foul types: screens are concentrated among the lowest calling scores and called less often. The pooled slope with foul-type controls added, or on the subsets without screens, is 0.22--0.34 (0.26 {[}0.05, 0.47{]} with play-type effects), positive, with an interval that excludes 0 except among the non-screen fouls by both definitions (0.22 {[}$-$0.01, 0.45{]}). Estimated separately within each type, the evidence is uneven: 0.45 {[}0.10, 0.77{]} within shooting fouls and 0.891 {[}0.193, 2.086{]} within screens, but 0.060 {[}$-$0.301, 0.387{]} within other personal fouls and 0.213 {[}$-$0.302, 0.756{]} within loose balls (supplement, Section S10.5; exploratory). The direction holds for the pooled slope, with a screen indicator, with play-type effects and without the report's screen plays; without the plays that either definition counts as screens the slope stays positive (0.22) but its interval reaches 0 ({[}$-$0.01, 0.45{]}); without the candidate screens alone the interval is {[}0.00, 0.45{]}. Two further checks are exploratory (supplement, Table S65). If screens by the segment's candidate flag are listed \(r\) times as completely as other fouls, the interval of the pooled slope reaches 0 only at \(r \approx 5.7\), that of the slope with play-type effects at \(r \approx 1.4\) (0.11 {[}$-$0.10, 0.32{]} at 2.73, R2's adjusted ratio in Section~\ref{sec:5.6}); by the report's play type neither reaches 0 up to \(r = 6\). Without the lowest eighth of the score the pooled slope is 0.27 {[}0.06, 0.50{]} and the slope with play-type effects 0.08 {[}$-$0.19, 0.37{]}; leaving out any other eighth changes the pooled slope little (0.38--0.43); with play-type effects the interval reaches 0 without eighth 3 ({[}$-$0.01, 0.45{]}). Against each departure taken alone the pooled slope holds up to a screen ratio of about 5.7 or a completeness factor of 2.25; acting together they bring its interval to 0 sooner: if screen candidates are listed more than about 1.65 times as completely as other fouls, a gradient inside the registered range (a factor of at most 1.67) suffices, and at the screen contrast's ratio of 2.73 a factor of 1.32; the point estimate stays positive throughout the registered range (supplement, Table S65). Within types the direction is suggested, not established. The factors below, 3.88 and 2.25, belong to the pooled registered model without type controls and do not carry over to the type-controlled slopes. With the foul type held fixed the corresponding factors are smaller (cancelling factor 2.5, interval reaching 0 at 1.2 with play-type effects), the latter inside the registered grid's range of 0.60--1.67. Under the linkage and retention weights, the end-game filters, month fixed effects, one call label and the registered sensitivity variants the slope lies between 0.345 and 0.463 (record-level re-estimates for the registered variants, segment-level fits otherwise), every interval excluding 0; the exception is the small subsample of 21 correct calls with an exactly observed whistle onset, 0.165 {[}$-$0.160, 0.550{]} (supplement, Tables S27 and S64; Sections S10.1, S10.2, S10.5 and S14.2). Figure~\ref{fig:2} shows the called share and the fitted propensity by eighth of the score, with the share of screens in each eighth. Over geometry, the propensity is below 0.6 for loose contact far from the basket and above 0.8 for tight contact near it, the same gradient as the floor's call rate, and partly a composition effect of foul types (supplement, Figure S11 and Section S10.5). Since the score is itself a model of calling, a positive slope is partly a validation of that model on a new label.

\emph{The direction.} The cancelling gradient is \(\gamma^{*} = -0.390\) per unit of the score. Among the 540 listed true fouls \(z\) runs from \(-4.86\) at the 10th percentile to \(-1.38\) at the 90th, a range of 3.47. Under the log-linear form of Section~\ref{sec:4.3}, \(\kappa(z) \propto \exp(\gamma z)\) (and likewise for relative retention), for the slope to be an artefact completeness would have to \emph{fall} by a factor of 3.88 across that range (1.67 per standard deviation), to about 0.26 at the 90th percentile if the 10th is listed completely. A relative retention of correct calls rising by 3.88 would do the same (Section~\ref{sec:4.2}). The two multiply: the slope depends on them only through the difference of their log-slopes, so any combination with product 3.88 cancels it (completeness halving with relative retention rising by 1.94, or the registered grid's largest ratio, 1.67, with retention rising by 2.32). With sampling error, the 95~\% interval reaches 0 at a combined factor of 2.25 (supplement, Figure S10): 3.88 cancels the point estimate, 2.25, from the lower end of the 2,000-resample interval, removes its significance. On the partially role-corrected data (Section~\ref{sec:7.1}), with its own score span (3.349), the factors are 3.61 and 1.98 (the latter from the 2,000-resample interval {[}0.203, 0.575{]}; supplement, Table S57); with completeness falling by 1.67, a rise in relative retention of 1.18 would then bring the interval to 0. Only the retention of correct calls can be checked; it falls slightly with the score (slope of being analysed on the score $-$0.33 {[}$-$0.70, 0.04{]}), the direction that understates the slope. Refitting slope and range in each of 1,000 game-cluster resamples gives a cancelling ratio of 3.88 {[}2.22, 7.45{]} (exploratory); 0.4~\% of resamples lie within the reach of the registered grid (at most 1.67) and 30~\% within that of the grid extended to 0.3 (at most 3.33); with a 3-df spline the ratio is 3.56 (Section S10.3). Given its sampling error, the sign of the pooled slope survives any log-linear completeness gradient the registered grid allows, taken alone, but not every one the extended grid allows; a completeness that differs by foul type is not of this form (Section~\ref{sec:4.1}), and combined with a screen ratio above about 1.65 a gradient inside the registered range already brings the interval to 0 (supplement, Table S65). These ranges are reporting devices, not estimates. A uniformly low completeness leaves the slope unchanged; only its fall across the score matters. If completeness is assumed to be at least some floor at every score, the fall is capped at 1/floor: 1.67 at the registered floor of 0.6, and 2.27 at a pointwise floor of 0.44, just beyond the 2.25 at which the slope's 95~\% interval reaches 0; the point estimate keeps its sign down to a pointwise floor of about 0.26 (factor 3.88). The pointwise floor is a sensitivity scenario, not an estimate: the within-season comparison (Section~\ref{sec:6}) suggests that overall completeness may be low, but an overall level near 0.44 does not bound completeness at every score.

\subsection{What the data say about information, by listing mechanism}\label{sec:5.5}

Whether the score carries information about fouling among all segments, hypothesis (b) of Section~\ref{sec:4.5}, depends on how missed fouls are listed (exploratory). \emph{If every missed foul is listed,} the prevalence is identified. On the segment population matched to the play set (24,393 segments, false calls set to zero), \(\pi(z) = P_W(D \mid z)/e(z)\) is about an order of magnitude higher at the 90th percentile of the score than at the 10th: 12.06 {[}9.79, 14.85{]} (2,000 resamples) under the linear logit, 19.7 {[}12.6, 35.4{]} under 3-df splines (Section S10.3), and 11.27 {[}9.39, 13.34{]} with every linked record of any retention status counted as listed, the label of Table~\ref{tab:4}. Counted directly by quintile of the score, the prevalence rises from 0.004 to 0.077 (supplement, Section S10.3). The calling probability rises at 0.91 per unit of the score and the propensity at 0.39, so under complete listing the gradient is mostly in the prevalence.

\emph{Under proportional selection,} missed fouls are listed with probabilities proportional to those of non-fouls at the same score: \(\kappa(z) = c\,q_{\text{nf}}(z)\), with \(q_{\text{nf}}\) the listing probability of uncalled non-fouls. The league's description of the listing motivates the scenario but does not imply it; the data give the joint listing and linkage of the label, proportionality is an added assumption, and the scale \(c\) is not identified. Solved to reproduce the observed listing of uncalled segments, the model implies the same numbers of listed missed calls and correct non-calls at every \(c\) (Section S8.4), and the scale decides the gradient. The slope of the propensity, by contrast, is identified under the scenario: about 1.0 (1.03 {[}0.89, 1.18{]}) at every scale up to the cap, and only the level depends on the unidentified scale (supplement, Table S33). We detect no clear difference in prevalence between the two percentiles up to \(\tilde c = 1/\max q = 2.15\), just below the joint cap onset \(\tilde c_0 = 2.27\) at which the listing of misses first reaches 1 on some segment (ratio 1.14 there): 0.86 {[}0.59, 1.24{]} at \(\tilde c = 1\), 1.08 {[}0.74, 1.54{]} at 2 and 1.11 {[}0.78, 1.61{]} at 2.15 (100 game-cluster resamples; the league reading of Table S32 gives close values with 2,000). At \(\tilde c = 1\) the scenario implies about 14 uncalled true fouls per game in the review scope, against 2.3 called; the share called reaches about 0.4 only by \(\tilde c = 5\). Across the seven months of reports, misses per correct call and correct non-calls per game rise together (Spearman 0.71, \(p\) = 0.07), as proportional selection would predict, but so would a changing grading standard or a changing listing of both kinds; the comparison does not test the scenario. Scales at which the listing of misses saturates at 1 for call-like contacts give a clearer positive difference (Table S31). The implied INC-versus-CNC AUC, 0.440 {[}0.377, 0.500{]} under every mechanism and scale (100 resamples; its interval reaching 0.5) (observed 0.467 on the segments of this label; 0.469 on the analysed records), says nothing the counts do not. Under every mechanism the fit implies 136.3 listed missed calls against 119 observed (+15~\%) and 1,080.1 correct non-calls against 1,097 (supplement, Section S8.4). A single label rule for segments with several records, and the league's own listing with in-frame linkage constant along the score, give close values: 11.24 {[}9.23, 13.67{]} and 0.781 under complete listing, 0.88--1.14 under proportional selection up to its cap onset \(c_0 = 2.18\) (Section S8.4 and Table S32). With every call listed, the data identify \(\pi(z)\,e(z)\); how it splits depends on a completeness scale the reports do not record (Section S8.5).

\begin{table}[!htbp]
\centering
\caption{Listed uncalled review-scope segments implied by the fitted models (exploratory), by listing of missed fouls: implied INC / CNC, prevalence ratio (\(\pi\) at the 90th over the 10th percentile of the score), share of true fouls called and implied INC-vs-CNC AUC. Population: the detected frame, the 24,393 segments of the population matched to the play set (508 called). Label: every uncalled segment linked to a listed non-call of any retention status (119 INC, 1,097 CNC; a segment with several records takes its first), listing and linkage jointly (Section~\ref{sec:4.2}). Parameters: complete joint listing (\(\kappa\,l_{\text{INC}} = 1\)), or proportional selection at the joint scale \(\tilde c = c\,l_{\text{INC}}/l_{\text{CNC}}\), which is not identified; without weights, false calls set to zero. Last row, a sensitivity: the 5 INC and 46 CNC in-frame candidates (on 5 and 45 distinct segments) added to the label (proportional selection at \(\tilde c = 1\), 2, \(1/\max q\): 0.87, 1.10, 1.15; dashes: not computed). The share called under complete listing (0.789) is the selection model's fit on this label, unweighted; the level on linked segments, and why segment levels lie above the play count, are in Section~\ref{sec:4.2}. Solver and linkage: Section S8.4; capped scales: Table S31; league reading: Table S32; unlinked rows: Section S10.2.}
\label{tab:4}
\footnotesize
\begin{tabular}{>{\raggedright\arraybackslash}p{0.26\linewidth} >{\raggedright\arraybackslash}p{0.12\linewidth} >{\raggedright\arraybackslash}p{0.12\linewidth} >{\raggedright\arraybackslash}p{0.10\linewidth} >{\raggedright\arraybackslash}p{0.10\linewidth} >{\raggedright\arraybackslash}p{0.10\linewidth}}
\toprule
Listing of missed fouls & Implied listed INC / CNC & Prevalence ratio, 90th / 10th percentile & Share of true fouls called (model-implied) & Uncalled true fouls per game & Implied INC-vs-CNC AUC \\
\midrule
Complete joint listing & 136.3 / 1,080.1 & 11.27 & 0.789 & 0.61 & 0.440 \\
Proportional selection, \(\tilde c = 1\) & 136.3 / 1,080.1 & 0.86 & 0.140 & 14.1 & 0.440 \\
Proportional selection, \(\tilde c = 2\) & 136.3 / 1,080.1 & 1.08 & 0.232 & 7.6 & 0.440 \\
Proportional selection, \(\tilde c = 1/\max q\) (2.15) & 136.3 / 1,080.1 & 1.11 & 0.244 & 7.1 & 0.440 \\
Sensitivity: label with the in-frame candidates (124 / 1,142 listed); complete joint listing & --- & 10.80 & --- & --- & 0.441 \\
\bottomrule
\end{tabular}
\end{table}

\subsection{Configuration contrasts and their tipping points}\label{sec:5.6}

\begin{figure}[!htbp]
\centering
\includegraphics[width=0.72\linewidth]{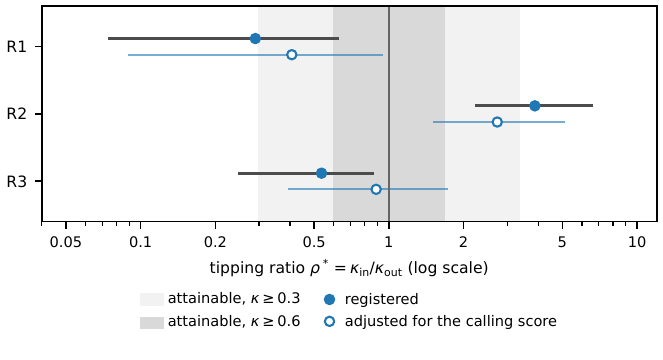}
\caption{Tipping points of the configuration contrasts. For each registered risk configuration (Section~\ref{sec:5.6}, Table~\ref{tab:5}), the relative completeness \(\rho^{*} = \kappa_{\text{in}}/\kappa_{\text{out}}\) that would cancel the observed difference in the calling propensity, on a logarithmic axis from 0.04 to 12, with 95~\% game-cluster bootstrap intervals: registered ratio from the raw called shares (filled; 300 resamples as registered, 2,000 in Table S24) and score-adjusted ratio (hollow; exploratory; 2,000 resamples). 1 marks no difference. Darker band: the range 0.60--1.67 attainable under the registered grid $\kappa$ $\in$ {[}0.6, 1{]}; lighter band: 0.30--3.33 with the grid extended to $\kappa$ = 0.3 (exploratory). Descriptive contrasts among listed true fouls. With end-game fouling situations excluded by the follow-up's filter, R2's adjusted ratio is 2.10, with the lower end of its interval at about 1 (Section~\ref{sec:5.6}). The share called over the registered grids of false-call rate and completeness is in Figure S9.}
\label{fig:3}
\end{figure}

Among the listed true fouls, screen fouls were called less often than others. The reports cannot separate this from a completeness that differs by foul type within the extended grid, a grading standard that differs by type, or which official decides, so we state it as a description. The three registered configurations, built from the rule predicates, are R1, a drive against a defender without legal guarding position and outside the restricted area; R2, screen contact with the screener moving off the defender's path; and R3, the ball close to the offensive player and rising, with no legal guarding position. Figure~\ref{fig:3} and Table~\ref{tab:5} give, for each, the share of listed true fouls called inside and outside it and the registered and score-adjusted tipping ratios of Section~\ref{sec:4.3}. R1 and R3 rest on 5 and 11 missed fouls inside. R2's difference is the most stable: its interval excludes 1 under every interval method and after score adjustment (2.73 {[}1.52, 5.08{]} adjusted); with end-game fouling situations set aside, the adjusted R2 ratio falls to 2.10; the lower end of its 95~\% interval lies at about 1 (0.94--1.02 across eight runs of 2,000 resamples; 1.05 by profile likelihood), so in that subpopulation the difference is not clearly separated from none (Section S10.5). R1's excludes 1 narrowly after adjustment (0.41 {[}0.14, 0.97{]}, profile likelihood) and R3's does not (0.89 {[}0.41, 1.77{]}, profile likelihood; the bootstrap intervals are in Table~\ref{tab:5}); by profile likelihood, a method not registered for it, even R3's registered interval reaches 1 (0.54 {[}0.26, 1.01{]}, against {[}0.25, 0.86{]} from the registered 300 resamples and {[}0.25, 0.93{]} from 2,000; supplement, Table S24). The intervals of the three ratios are not adjusted for multiplicity. Inside R1 and R3 true fouls were called \emph{more} often than outside; for R2 the difference is reversed and the largest. At equal score, missed fouls on such screens would have to be listed 2.73 times as completely as other missed fouls to remove R2's point difference: outside the registered range 0.60--1.67, inside the extended range 0.30--3.33. A relative completeness inside the registered range suffices to remove its significance, since the lower end, 1.52, lies below 1.67. On the partially role-corrected data (Section~\ref{sec:7.1}) the adjusted ratio is 2.41 {[}1.29, 4.52{]}, with the same reading; three missed screen fouls leave R2, and its registered ratio falls from 3.87 to 3.34 {[}1.89, 6.11{]}. The tipping ratios also need equal relative retention (Section~\ref{sec:4.2}), which screens do not have; for screens defined by the reports' text, retention understates the contrast rather than producing it (Section S10.2).

\begin{table}[!htbp]
\centering
\caption{Configuration contrasts: the registered enrichment test (SA4; INC against CNC, Fisher's exact test with Holm's correction, records treated as independent), the registered called shares and tipping ratios (300-resample game-cluster intervals), and the exploratory score-adjusted tipping ratio \(\rho^{*} = \exp(-\theta)\) (95~\% CI). With end-game fouling situations set aside under the follow-up's filter, the adjusted ratios are 0.34 {[}0.00, 0.88{]} (resampled interval degenerate at 0; profile likelihood {[}0.08, 0.99{]}), 2.10 {[}1.01, 4.28{]} and 0.86 {[}0.32, 1.95{]} (2,000 resamples; for R2 the lower end is 0.94--1.02 across eight seeds; Section S10.5). Records for SA4 (R1: 6 INC, 84 CNC inside); listed true-foul segments for the shares and ratios (435 CC, 105 INC; R1: 64 / 5). A game-clustered check of SA4 is in Section S7.3.}
\label{tab:5}
\footnotesize
\begin{tabular}{>{\raggedright\arraybackslash}p{0.21\linewidth} >{\raggedright\arraybackslash}p{0.12\linewidth} >{\raggedright\arraybackslash}p{0.10\linewidth} >{\raggedright\arraybackslash}p{0.08\linewidth} >{\raggedright\arraybackslash}p{0.15\linewidth} >{\raggedright\arraybackslash}p{0.17\linewidth}}
\toprule
Configuration & SA4 odds ratio (p; Holm) & CC share inside / outside & CC / INC inside & \(\rho^{*}\) registered {[}95~\% CI{]} & \(\rho^{*}\) score-adjusted {[}95~\% CI{]} (exploratory) \\
\midrule
R1: drive, no legal guarding position, outside the restricted area & 0.71 (0.565; 1.000) & 0.928 / 0.788 & 64 / 5 & 0.29 {[}0.08, 0.62{]} & 0.41 {[}0.09, 0.94{]} \\
R2: screen contact, screener moving off the defender's path & 1.37 (0.178; 0.534) & 0.574 / 0.839 & 39 / 29 & 3.87 {[}2.25, 6.54{]} & 2.73 {[}1.52, 5.08{]} \\
R3: ball close to the offensive player and rising, no legal guarding position & 0.80 (0.644; 1.000) & 0.876 / 0.792 & 78 / 11 & 0.54 {[}0.25, 0.86{]} & 0.89 {[}0.40, 1.72{]} \\
Any screen candidate --- the data-preparation layer's flag on which R2 is built (exploratory) & --- & --- & 60 / 41 & --- & 2.93 {[}1.72, 5.06{]} \\
\bottomrule
\end{tabular}
\end{table}

\subsection{Other registered analyses}\label{sec:5.7}

The other registered analyses are in the supplement. SA2 and SA5 are reported in Section~\ref{sec:5.3} and SA4 in Table~\ref{tab:5}; SA1 and SA3 are consistent with the main test (Table S23; SA3: supplement, Section S7.2 and Figure S5). Of the negative controls, N2 and N4 behaved as registered; N1 was non-informative, and N3, registered to fall clearly below the main test, did not (0.518 {[}0.456, 0.585{]}), since the main test is itself near one half; so the controls cannot say whether the main test reflects a lack of information or a failure of method (Section S7.5). In the review budget the calling score does not say which assessments were errors (Section S7.8); the sensitivity variants are in Section S7.9 and Table S27, the extrapolation of the levels to the rest of the game under A3 in Section S7.7, and the home-court example, with no detectable difference, in Section S7.4.

The registered follow-up (Section S9; registered after the first-stage results) asked whether segments in the last two minutes of close games are called at a different rate from segments with an equal geometric score elsewhere, and how calls on screens relate to a centroid reading of the screen rule. It found higher odds of a call for \emph{close-late} segments (the fourth period or overtime, at most 2:00 left, a margin of at most 5 points at the 2:00 mark) at equal geometric score: odds ratio 1.331 {[}1.218, 1.454{]}, ``more calls'' by the registered rule. The crude call rates disclosed in the registration already implied an odds ratio of about 1.34, a point the registration did not state; the registered estimate adds an interval and a geometric adjustment rather than a new direction. With end-game fouling situations excluded by a post-hoc filter, the exploratory analysis found no detectable difference in the remaining sample (1.008 {[}0.898, 1.130{]}), while late segments of games that are not close keep a small excess (1.110 {[}1.014, 1.216{]}). The two estimates describe different populations; the filtered one does not replace the registered result or establish equal calling rates. Neither separates foul prevalence from the propensity to call a true foul, so neither identifies transport (A3). On screens, offensive fouls are called on very few candidate screens, and the call rate rises slightly with the screener's off-path speed (registered slope +0.051 {[}+0.019, +0.082{]} per ft/s), on a permissive centroid proxy of the screen rule (Section S9.4).

\section{Selection, linkage and retention: what each result needs}\label{sec:6}

Table~\ref{tab:6} summarises the principal empirical results, their grade and the premises each needs (Section~\ref{sec:4.1}); the checks are in Sections 3--5. For completeness, what matters is its shape more than its level (Sections~\ref{sec:5.4} and~\ref{sec:5.5}); for retention, only that of correct calls can be checked along the score (Section~\ref{sec:4.2}). Two points add to the table.

\emph{Linkage.} The label of the segment models is listing and linkage jointly (Section~\ref{sec:4.2}). Of the listed non-calls, 45 INC and 333 CNC link to no segment, mostly because no segment of the pair lies near the reported clock (24, 132) or the play falls in a tracking hole (12, 117); over all listed rows the linkage rates are 0.735 for INC and 0.774 for CNC. For listed non-calls whose pair has a segment near the reported clock they are 0.954 and 0.952, with slopes on the score of 0.45 {[}$-$0.69, 1.41{]} and 0.51 {[}0.30, 0.76{]}: there the data do not contradict the condition for reading the label as the league's listing, though the slope for missed calls is imprecise; outside that range linkage cannot be studied along the score (supplement, Section S10.2).

\emph{The listing within the season.} The 223 intersection games list 0.76 missed calls per game, with a called share CC/(CC + INC) of 0.777; the season's other 187 report games list 1.26 per game with a called share of 0.693, a gap of the calendar rather than of tracking coverage. Listed missed calls per correct call rose after January while correct calls per game showed no trend (supplement, Section S14.2); a more inclusive listing, a stricter grading or a change in calling fit the rise, and the reports cannot separate them. Per correct call, the intersection games list 0.597 {[}0.447, 0.765{]} times the missed calls of February to April and 0.438 {[}0.321, 0.625{]} times those of March. These ratios are relative completeness: they are the completeness of the intersection games only if neither calling nor grading changed and the later listing was complete (supplement, Section S10.1). Month fixed effects leave the slope and the share called unchanged (Section S14.2). The listing of correct non-calls grew about 2.5-fold within the analysed window, which the main test and SA1--SA4 pool.

\begin{table}[!htbp]
\centering
\caption{The principal empirical results: grade (Section~\ref{sec:1}) and premises (Section~\ref{sec:4.1}). Rows describe the review scope of the October--January 2015-16 reports, except the first (the reports alone), the close-late row (all 631 tracked games) and the last (the reports of October to April).}
\label{tab:6}
\footnotesize
\begin{tabular}{>{\raggedright\arraybackslash}p{0.42\linewidth} >{\raggedright\arraybackslash}p{0.16\linewidth} >{\raggedright\arraybackslash}p{0.21\linewidth} >{\raggedright\arraybackslash}p{0.10\linewidth}}
\toprule
Result & Grade & What it needs & Where \\
\midrule
Share of true-foul plays called, listed or not: at most 0.777 {[}0.744, 0.806{]} (0.777 {[}0.745, 0.807{]} with the 3 unassessed calls counted as correct), exact under complete listing; missed true fouls at least 0.76 {[}0.64, 0.90{]} per game; no lower bound of practical use & Exploratory & Every call listed; take fouls removed & Section~\ref{sec:3} \\
Listed missed fouls not strongly separated from listed correct non-calls: AUC 0.469 {[}0.415, 0.526{]}, ``insufficient''; SA5 0.471 {[}0.411, 0.534{]} & Confirmatory & The registered rule; a statement about ranking among the listed plays & Sections 4.5, 5.3 \\
Other registered analyses: SA1, SA3, the home-court example and the review budget consistent with the main test; N2 and N4 as registered, N1 non-informative, N3 not below the main test & Registered, descriptive & As registered & Section~\ref{sec:5.7} \\
The propensity rises with the calling score: pooled slope 0.390 {[}0.220, 0.577{]}; 0.22--0.34 with type controls or without screens; within types suggested, not established (uneven within single types; 0.08 {[}$-$0.19, 0.37{]} with play-type effects without the lowest eighth) & Registered, descriptive; type analyses exploratory & A5; completeness not falling with the score; a retention ratio not rising with it & Section~\ref{sec:5.4} \\
Factors that cancel the slope or bring its interval to 0 under log-linear completeness and retention: pooled 3.88 {[}2.22, 7.45{]} (1,000 resamples) and 2.25; with play-type effects 2.47 and 1.17, and the interval reaches 0 if screens (by the candidate flag) are listed about 1.4 times as completely as other fouls, or without the lowest eighth of the score; screen ratio and gradient together: interval reaches 0 from a ratio of about 1.65 within the registered range; point estimate positive throughout & Exploratory & The log-linear form; a pointwise floor is a scenario, not an estimate & Section~\ref{sec:5.4} \\
Share of true fouls called on detected segments, by a model: 0.784 {[}0.745, 0.820{]} (linked segments matched to the report plays, weighted for in-frame retention); 0.791 {[}0.749, 0.825{]} (registered record); both above 0.777, the first for reasons of population, the second also through the lower retention of missed calls & Registered record; linked segments: exploratory & A1 (at \(\kappa = 1\)); the single-index form; A5 & Sections 3, 4.2 \\
Prevalence ratio, 90th to 10th percentile of the score: 12.1 (linear logit) and 19.7 (splines) under complete listing; no clear difference under proportional selection up to the cap onset & Exploratory & The listing mechanism; only the product of prevalence and propensity is identified & Section~\ref{sec:5.5} \\
Configuration contrasts R1--R3: screen fouls (R2) called less often; adjusted tipping ratio 2.73 {[}1.52, 5.08{]}, lower end about 1 without end-game fouling situations & Registered, descriptive; R2's reading exploratory & The ratio of completeness between the groups; equal relative retention & Section~\ref{sec:5.6} \\
Close-late call-rate contrast at equal geometric score: OR 1.331 {[}1.218, 1.454{]}, a direction the disclosed crude rates implied; 1.008 {[}0.898, 1.130{]} without end-game fouling situations & Registered, descriptive (second registration); filtered: exploratory & A comparison of calling rates, uninformative about A3 & Section~\ref{sec:5.7} \\
Listed missed calls per correct call rose within 2015-16; per correct call the October--January reports list 0.597 times the missed calls of February--April & Exploratory & Calling and grading unchanged, to read the ratio as relative completeness & Section~\ref{sec:6} \\
\bottomrule
\end{tabular}
\end{table}

\section{Discussion and conclusion}\label{sec:7}

\subsection{Limitations}\label{sec:7.1}

\textbf{What two-dimensional centroids cannot see.} Arms, hands, feet, torso orientation, verticality and jump height are unobserved, and every rule predicate is a proxy (Section~\ref{sec:5.1}; supplement, Sections S3.4 and S6.4). Positional jitter is small, so the dominant error is structural (supplement, Section S11).

\textbf{Detection, timing and the model.} The detector misses some called fouls, the whistle onset is mostly estimated, and the main score is, by design, a readable \(L_2\) logistic regression without player or official effects; the registered variants, SA5 with gradient boosting and the exploratory checks give the same answers (Section~\ref{sec:5.3}; supplement, Sections S3, S4, S10.2 and S10.5).

\textbf{Offence and defence roles.} The possession ledger of the data-preparation layer can reverse offence and defence near possession starts, and a post-hoc correction there is partial. With every feature rebuilt on the partially corrected roles, or with the segments within 3 s of a possession start removed, no reading changes (supplement, Section S10.2).

\textbf{Links of the listed non-calls.} A listed non-call is linked to a segment by its clock window and the players' closest approach (Section~\ref{sec:2.3}); whether the linked segment is the contact the report describes was not checked record by record. The segment-level results hold for these algorithmic links: the checks of Section~\ref{sec:6} and the supplement (Section S10.2) show how sensitive they are to the errors examined there, not how often a link is right. The play-level count bound does not depend on linkage.

\textbf{Review scope and season.} Every statement joining the tracking to the reports concerns the review scope of the October--January 2015-16 reports (for the calling model, all 631 tracked games and the whole game; for months and seasons, the reports alone). Transport to the rest of the game needs A3 (supplement, Section S7.7), which none of our analyses tests. The analysed records come from 221 of the 223 intersection games and include 109 listed missed calls, against the 406 contact-foul misses in the season's reports. On the reports' text alone, 2015-16 resembles later seasons in its \emph{calls} (a raw CC/(CC + INC) of 0.73, inside the 0.66--0.87 of the later seasons up to 2024-25) but not in its \emph{listed non-calls}, whose listing grew later, so no result that conditions on the correct non-calls, including the main test, should be transported (supplement, Section S14).

\subsection{What would change the conclusions}\label{sec:7.2}

Four things would. The first is an independent video review of the 22,669 uncalled review-scope segments that the reports do not list. The league lists the more noticeable non-calls and a panel watching broadcast video also detects noticeable fouls more readily, so the two lists are dependent through a salience that cannot be stratified away: the panel would overstate completeness and give a tighter upper bound than 0.777, not an estimate. Broadcast footage also shows the score, the teams and the crowd, so the panel cannot be blind to them without edited clips (frame, sample sizes and calibration: supplement, Section S10.6). The second is features that see limbs (pose estimates, or the richer tracking the league now collects), which would make the rule predicates less of a proxy and let the main question be asked of segments no reviewer selected. The third is a registered analysis of a later season with the calling model transported, separating changes in the listing (Section~\ref{sec:6}) from changes in calling. The fourth needs no tracking: did the listed missed calls rise within the categories newly brought in when later reports expanded their listing of non-calls? If not, the marginal misses were already listed; the comparison is informative only if the true miss rate and the calling standard stayed stable, and only for the materiality layer of completeness.

\subsection{Beyond the last two minutes}\label{sec:7.3}

The structure studied here, with every decision logged, a selective review in some contexts, and narratives written after the verdict, is not peculiar to basketball (video review in other sports, appeals, clinical audits). The identification argument transfers with its condition: a selective review that covers every positive decision identifies the propensity of a true positive to be decided positively, provided it lists every missed positive. So do the reporting devices (a one-sided bound for levels, a relative completeness for differences, a completeness gradient for slopes) and the lesson of Section~\ref{sec:4.5} on tests inside a sample whose negatives the reviewers chose. So does a lesson of the design: pairing an official review with sensor data lets the assessment leak into a feature, a usability rule or a sampling frame and create a signal that out-of-sample evaluation confirms, and none of the eight pitfalls we met --- among them a usability rule that needed tracked frames after the contact, smoothing that reads past the end of the window, and a review text that depends on the assessment --- was visible in a cross-validated performance number. Of every filter, anchor and segment rule one must ask whether it could be computed had the decision gone the other way (supplement, Section S4 and Table S11).

\subsection{Conclusion}\label{sec:7.4}

Under a completeness condition it does not itself guarantee, a selective review of officiating decisions identifies how the propensity to call a true foul varies with a calling score built from the geometry of the contact; it does not identify the completeness of its own listing or the propensity outside the review scope, which we report as bounds and tipping points.

In the only season with public player tracking, to our knowledge, at most about 78~\% of all true-foul plays, listed or not, in the review scope of the 223 games of October 2015 to January 2016 with tracking were called; this bound uses only the report counts, and they give no lower bound of practical use. The propensity rises with the calling score, partly between foul types, if completeness does not fall with the score and the relative retention of correct and missed calls does not rise with it; within types the rise is suggested, not established, since with play-type effects its interval reaches 0 under a completeness gradient of 1.17, if screens (by the candidate flag) are listed about 1.4 times as completely as other fouls, or without the lowest eighth of the score. By the pre-registered criterion the calling score showed no strong directional separation of the listed missed fouls from the listed correct non-calls; the test had limited power to declare the score informative, but its interval excludes strong separation in either direction (Section~\ref{sec:5.3}), and it concerns ranking among the listed plays, not the score's information on fouling.

\bibliography{refs}

\appendix
\setcounter{table}{0}
\renewcommand{\thetable}{\thesection\arabic{table}}
\makeatletter\@addtoreset{table}{section}\makeatother
\section{The two registrations and the deviation log}\label{app:A}

\textbf{First registration.} We registered the analysis plan (Section~\ref{sec:5}) on the Open Science Framework on 2026-10-01 at 15:15:23 UTC (osf.io/sr243; OSF Preregistration template), archiving the registration document, the three lists of play identifiers (confirmatory, sensitivity and pilot) with their SHA-256 hashes, and the frozen analysis code \texttt{analysis\_\allowbreak{}code.tar.gz}, unchanged when the results were recorded; the archived files reproduce the registered hashes (supplement, Section S13.2). We wrote the plan with prior access to the data, which the registered document discloses item by item and the supplement summarises (Section S7.1). The registered analyses used the archived code, and the deviation log records no deviation for this plan. The selection comparison (supplement, Section S8) and every analysis added after registration are labelled exploratory: additions, not deviations. One reading changed after registration: the plan read the main test as bearing on whether the score carries information about fouling, and we read it as a statement about ranking among the listed plays (Sections~\ref{sec:1} and~\ref{sec:4.5}).

\textbf{Second registration (the follow-up).} The follow-up's plan (Section~\ref{sec:5.7}; supplement, Section S9) was registered on 2026-10-01 at 17:25:30 UTC (osf.io/hnzdc), after every result of the first plan was known and before any follow-up analysis was run. One reporting deviation is recorded: the registered text states that the frozen code was uploaded as an archive, but it was not. The registered text records the SHA-256 of that archive, \texttt{analysis\_\allowbreak{}code\_\allowbreak{}stage2.tar.gz}, which the archive rebuilt after registration reproduces, so the omission affects where the code can be obtained, not whether it is the code that ran (supplement, Section S13.2). The split by score status and the end-game-situation filter were added afterwards and are exploratory. The registered transport grid \(\delta \in \{-0.5, 0, 0.5\}\) is symmetric; we define \(\delta = \operatorname{logit} e_{\text{in}} - \operatorname{logit} e_{\text{out}}\), as the code implements it, so the registered envelope is unchanged. The follow-up's registered reading of its primary estimate is ``more calls''; Section~\ref{sec:5.7} reports it, notes that the crude call rates disclosed in the registered text already implied its direction (supplement, Section S9.1), and adds an exploratory filtered analysis that does not replace it. The second registration's text quotes the call model's discrimination as 0.846 overall, the value before the fix of pitfall 7 (supplement, Section S4.7); the fixed model gives 0.841, and no analysis is affected. The registered first-stage analyses ran after the first registration (15:15 UTC) and their results were committed at 15:33 UTC; the second plan, its code and a dry run on permuted outcomes were written and frozen between 16:28 and 17:13 UTC, with the first-stage results known, and registered at 17:25 UTC.

\textbf{Post-hoc observations on registered ranges.} That the registered false-call rates of 0.01--0.05 are incompatible with the 47 listed incorrect calls (Section~\ref{sec:4.2}) was observed after the registered analyses. The registered envelope of the share called, {[}0.695, 0.886{]} (uncalled true fouls per game {[}0.090, 1.219{]}), is archived here and in Table S47 as the registered record; its upper end comes from the registered false-call rates of 0.01--0.05, which the listed calls rule out (Section~\ref{sec:4.2}); nothing registered was recomputed. With false calls anchored on the listed calls, the registered computation on the 24,522 bounds units (the review-scope segments, including the 120 called segments the reports list as take or other fouls; 222 games, 617 called segments) gives a share called of 0.694--0.790 over \(\kappa \in [0.6, 1]\) (0.640--0.818 with sampling) and 0.682--1.134 uncalled true fouls per game (0.575--1.407); on the population of the count bound the same range is 0.68--0.78 and 0.76--1.27 per game (Section~\ref{sec:3}).

Both registrations are under embargo until 2028-10-01. The journal's referees can read them through anonymised view-only links.

\section{Data and code availability, licence and ethics}\label{app:B}

\textbf{The review package.} An anonymous review package has been made available to the journal's referees. It contains the derived tables (without tracking coordinates or referee identifiers), the analysis code with its lock file, and a provenance file that maps every table and figure citing a computed result to the script, output and seed behind it; steps that need the tracking frames, the report text or the public report table are included as code only (supplement, Section S13.2). The package also holds the frozen code archive of the second registration, unchanged so that its SHA-256 matches the registered value. The data-preparation package, which builds the derived event tables and is pinned at a fixed commit, is the authors' own; its name appears in the frozen code of both registrations and of the review package's stage-2 archive, it is not part of the review package, and its repository and citation will be given on acceptance.

\textbf{The release package.} On acceptance, the released derived data will form one folder (98 MB) with its own README and a manifest, published with its repository and archive identifiers. It differs from the review package in its file layout, its pseudonymised player fields and its reduced columns, and contains no raw tracking coordinates, no official's name, identifier or token, and no comment text or player names. The supplement lists the files, the exclusions and the steps to rebuild everything from the public sources (Section S13.1), and the provenance file, with the result files, scripts and seeds of each table and figure, and the lock file (Section S13.2).

\textbf{Licence and credit.} The table of reports is taken from the community repository \texttt{atlhawksfanatic/\allowbreak{}L2M} \citep{atlhawksfanatic2025} under the MIT License; the NBA publishes the reports, and the play-by-play is credited to NBA.com. Redistribution and officials: supplement, Section S13.3. The derived tables' licence, a takedown contact, a statement on consultation with the league and archive identifiers follow on acceptance.

\textbf{Ethics and statements.} The study uses public data only, infers no player's intent, and reports no official-level result (supplement, Section S13.3). No ethics review was required. Informed consent: not applicable (public data). Research funding: none. Conflict of interest: none. Author contributions: T.G., conceptualization, methodology, software, validation, formal analysis, writing (original draft), writing (review and editing) and project administration; Y.Z., conceptualization, methodology, software, validation, data curation and writing (review and editing).

\clearpage
\includepdf[pages=-]{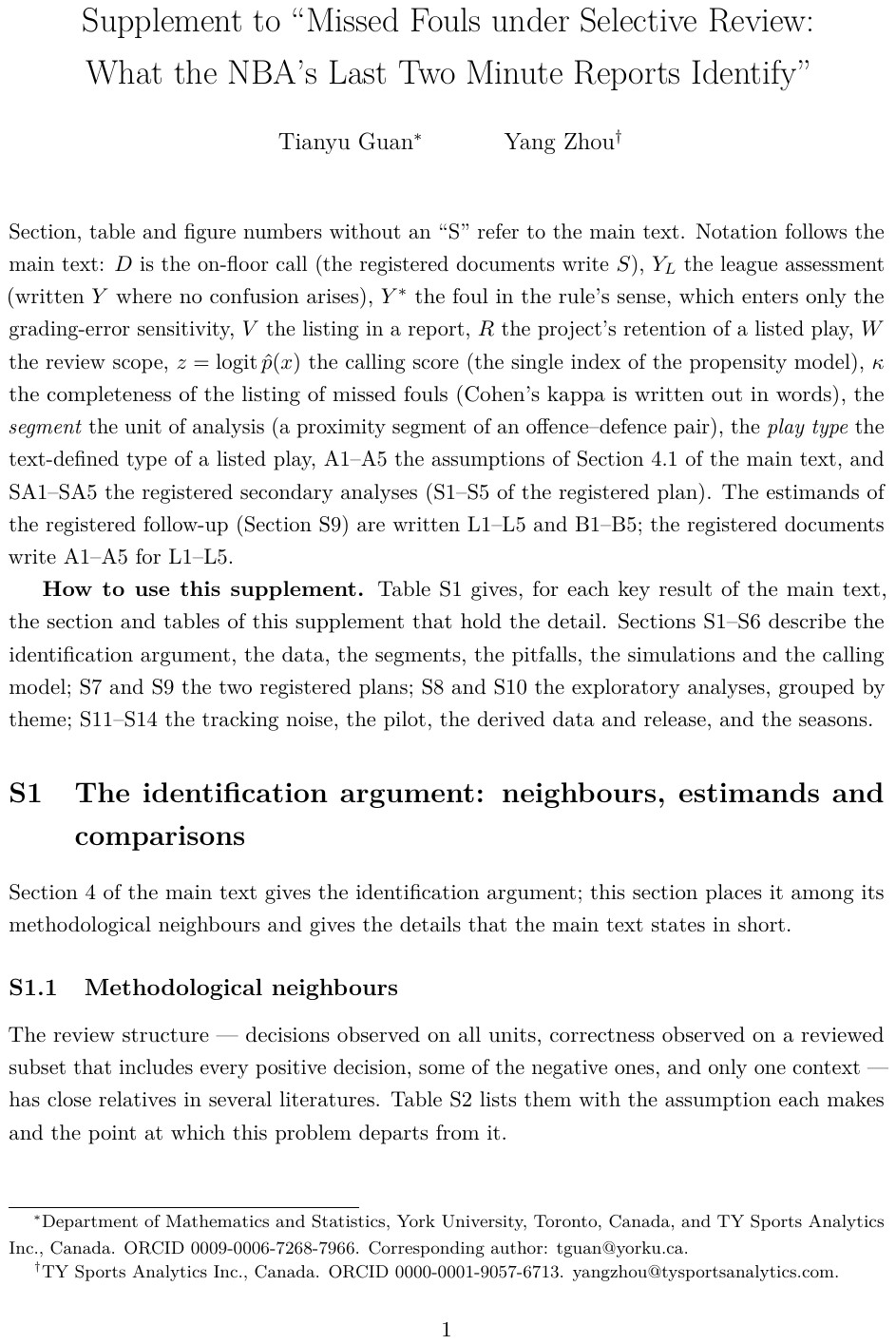}

\end{document}